\documentclass[letter,oldversion]{aa}
\usepackage{graphicx, natbib, lscape, amssymb}
\usepackage[english]{babel}
\newcommand{\gppr}{\stackrel{>}{\scriptstyle \sim}}
\newcommand{\gappr}{\raisebox{-0.4ex}{$\gppr$}}
\newcommand{\lppr}{\stackrel{<}{\scriptstyle \sim}}

\newcommand{\Mwd}{\mbox{$M_\mathrm{WD}$}}
\newcommand{\Msec}{\mbox{$M_\mathrm{sec}$}}
\newcommand{\Msun}{\mbox{$M_{\odot}$}}

\newcommand{\Porb}{\mbox{$P_\mathrm{orb}$}}

\begin{document} 

\title{The evolution of the self-lensing binary KOI-3278:\\ evidence of extra energy sources during CE evolution}
\titlerunning{The evolution of KOI-3278}
\author{M. Zorotovic\inst{1}, M.R. Schreiber\inst{1,2}, S.G. Parsons\inst{1}}
\authorrunning{M. Zorotovic et al.}
\institute{Instituto de F\'isica y Astronom\'ia, Universidad de Valpara\'iso, Av. Gran Breta\~na 1111, Valpara\'iso, Chile\\
\email{mzorotov@dfa.uv.cl}
\and Millennium Nucleus ``Protoplanetary Disks in ALMA Early Science'', Universidad de Valpara\'iso, Casilla 36-D, Santiago, Chile }
\offprints{M. Zorotovic}
\date{Received: 19 June 2014/ Accepted: 21 July 2014}

\abstract{Post-common-envelope binaries (PCEBs) have been frequently used to observationally constrain models of close-compact-binary evolution, 
in particular common-envelope (CE) evolution. However, recent surveys have detected PCEBs consisting of a white dwarf (WD) exclusively with 
an M dwarf companion. Thus, we have been essentially blind with respect to PCEBs with more massive companions. 
Recently, the second PCEB consisting of a WD and a G-type companion, the spectacularly self-lensing binary KOI-3278, has been identified.
This system is different from typical PCEBs not only because of the G-type companion, but also because of its long 
orbital period. Here we investigate whether the existence of KOI-3278 provides new observational constraints on theories of CE evolution.
We reconstruct its evolutionary history and predict its future using BSE, clarifying the proper use of the binding 
energy parameter in this code.
We find that a small amount of recombination energy, or any other source of extra energy, is required to reconstruct the evolutionary 
history of KOI-3278. Using BSE we derive progenitor system parameters of $M_{1,i} = 2.450\Msun$, $M_{2,i} = 1.034\Msun$, and 
$P_{orb,i} \sim 1300\,d$. We also find that in $\sim9\,Gyr$ the system will go through a second CE phase leaving behind a double 
WD, consisting of a C/O WD and a He WD with masses of $0.636\Msun$ and $0.332\Msun$, respectively.
After IK\,Peg, KOI-3278 is the second PCEB that clearly requires an extra source of energy, beyond that of orbital energy, to contribute to the 
CE ejection. Both systems are special in that they have long orbital periods and massive 
secondaries. This may also 
indicate that the CE efficiency increases with secondary mass.}


\maketitle

\section{Introduction}

Post-common-envelope binaries (PCEBs) consisting of a white dwarf (WD) and a main-sequence (MS) companion are close-binary stars with orbital 
periods typically shorter than a day. Their discovery \citep{kraft58-1} immediately raised the question of their 
origin, because the progenitor of the WD must have been much bigger than the separation of the two stars in the currently observed system. 
Based on the pioneering works of \citet{paczynski76-1} and \citet{webbink84-1} the puzzle now seems to be solved. 
If the initially more massive star fills its Roche lobe as a giant and if the mass ratio $q=M_{\mathrm{donor}}/M_{\mathrm{gainer}}$ exceeds 
a critical value $q_{crit}$\footnote{Early calculations led to $q_{crit}\sim1$ \citep{webbink88-1} while more recent works do not 
exclude values as high as $q_{crit}=1.5$ \citep{passyetal12-1,woods+ivanova11-1}.}, dynamically unstable mass transfer is generated. 
This leads to the formation of a common envelope (CE) engulfing both the core of the primary (the future WD) and the secondary star. 
The CE is expelled at the expense of orbital energy and angular momentum leaving behind a short period PCEB consisting of the compact core of the primary 
and the secondary star. 

Despite significant recent progress \citep{ricker+taam12-1},
numerical calculations still fail to simultaneously cover the large range of 
time and spatial scales involved in CE evolution and to make detailed
predictions for the parameters of the emerging PCEB. Therefore, a simple energy 
equation relating the binding energy of the envelope to the change in orbital 
energy parametrized with the so-called CE efficiency ($\alpha_{\mathrm{CE}}$) is 
normally used to predict the outcome of CE evolution. Such an approach
requires observational constraints on the efficiency parameter.  

Recent surveys of PCEBs have established large samples of close binaries 
containing a WD and an M-dwarf companion \citep[e.g.,][]{nebot-gomez-moranetal11-1,parsonsetal13-1}.
These samples have been proved useful to understand several aspects of 
close-compact-binary evolution \citep[e.g.,][]{zorotovicetal10-1}; however, 
they only contain low-mass secondary stars.   
The predicted significant population of PCEBs containing a WD plus a 
massive ($\gappr1\Msun$) secondary star \citep[see, e.g.,][]{zorotovicetal14-1} 
has not yet been identified. 
This is because such a massive MS star completely outshines the WD 
at all wavelengths longer than UV. 
Finding and analyzing the evolutionary history of these PCEBs is crucial 
not only because the CE efficiency may depend
on the mass of the secondary, as speculated by
e.g., \citet{politano+weiler07-1}, but also 
because these systems may 
hold the key to understanding one of the oldest problems in astrophysics: 
the progenitor problem for supernovae Type\,Ia 
\citep[SN\,Ia, see, e.g.,][ for a recent review]{wang+han12-1}. 
PCEBs with a massive MS secondary star 
are the progenitors of the two most popular channels proposed towards SN\,Ia. 
In the single-degenerate channel \citep{whelan+Iben73-1}
these PCEBs start thermal-timescale 
mass transfer which allows the WD mass to grow until it eventually explodes 
as a SN\,Ia. In the double-degenerate channel \citep{webbink84-1}
a WD with a close and massive companion (either a PCEB or a close binary emerging from stable mass transfer)
evolves into a CE phase which leaves behind a double-degenerate 
system that may finally merge and, in the case of two C/O WDs with a total mass
exceeding the Chandrasekhar limit, produce a SN\,Ia explosion. 

Recently, the second WD with a close and massive companion star has been identified
(after IK\,Peg). This system, KOI-3278, identified using data from the Kepler
spacecraft \citep{kruse+agol14-1}, is remarkable not only for its long orbital
period (88.18 days), but also because it is eclipsing. This combination of long
period and high inclination results in a spectacular five-hour pulse once every 
orbit, caused by the 0.634\Msun\,C/O WD acting as a gravitational lens
as it passes in front of its 1.042\Msun\,MS companion. 

Given the small separation and the masses of the two stars, KOI-3278 must have evolved through a CE. Assuming 
that stable mass transfer occurred in KOI-3278 would require a large critical mass ratio ($q\sim1.5$) {\em{and}} strong 
wind mass loss of the WD progenitor on the AGB. While this configuration cannot be completely excluded, the resulting 
stable mass transfer could not have reduced the binary orbital period to the measured 88.18 days. Thus, KOI-3278 is the
second PCEB containing a massive companion star.

Here we reconstruct the evolutionary history of KOI-3278 to derive constraints on the CE efficiency and predict its future to evaluate
whether it might be the first progenitor of a double WD that will be formed through two CE phases. 

\section{Constraints on CE evolution from KOI-3278}

In its simplest form, the energy equation describing CE evolution can be expressed as
\begin{equation}\label{eq:alpha}
E_\mathrm{bind} = \alpha_{\mathrm{CE}}\Delta E_\mathrm{orb}.
\end{equation}
The most basic assumption is to approximate the binding energy only by the gravitational energy of the envelope,
\begin{equation}\label{eq:Egr}
E_\mathrm{bind} = E_\mathrm{gr}=-\frac{G M_\mathrm{1} M_\mathrm{1,e}}{\lambda R_\mathrm{1}},
\end{equation}
where $M_\mathrm{1}$, $M_\mathrm{1,e}$, and $R_\mathrm{1}$ are the total mass, envelope mass, and radius of the primary star, 
and $\lambda$ is a binding energy parameter that depends on the structure of the primary star. 
Simulations of PCEBs \citep[e.g.,][]{dekool+ritter93-1,willems+kolb04-1,politano+weiler06-1} 
have been performed assuming different values of $\alpha_{\mathrm{CE}}$ and assuming $\lambda = 0.5$ or 
$1.0$. However, keeping $\lambda$ constant is not a very realistic assumption for all types of possible primaries 
as was pointed out by e.g., \citet{dewi+tauris00-1}. Very loosely bound envelopes in more evolved stars
can reach much higher values, especially if the 
recombination energy $U_\mathrm{rec}$ available within the envelope supports the ejection process. 
Therefore, a more realistic form for the binding energy equation is
\begin{equation}\label{eq:Eball}
E_\mathrm{bind}=\int_{M_\mathrm{1,c}}^{M_\mathrm{1}}-\frac{G m}{r(m)}dm + \alpha_{\mathrm{rec}}\int_{M_\mathrm{1,c}}^{M_\mathrm{1}}U_\mathrm{rec}(m),
\end{equation}
where $\alpha_\mathrm{rec}$ is the efficiency of using recombination energy, i.e., the fraction of recombination energy that contributes 
to the ejection process. The effects of the extra energy source can be included in the $\lambda$ parameter by equating 
Eqs.\,\ref{eq:Egr} and\,\ref{eq:Eball}.

\subsection{Previous observational constraints on CE efficiencies}

While the above straightforward energy equation accurately describes the basic idea 
of CE evolution, it requires observational constraints 
to estimate the efficiencies. Several attempts to provide 
such constraints have been made using PCEBs consisting of a 
WD and a late-type (M dwarf) companion. 
In \citet{zorotovicetal10-1}
we have shown that the evolutionary history of the identified PCEBs 
can be reconstructed assuming that both efficiencies are 
in the range of $0.2-0.3$. The case for such 
relatively small efficiencies has recently been strengthened 
by \citet{toonen+nelemans13-1} and \citet{camachoetal2014-1} who performed 
binary population models of PCEBs taking into account selection effects 
affecting the observed samples. 
While the relative contributions of recombination and orbital energy 
remain unclear \citep[e.g.,][]{rebassa-mansergasetal12-1}, the small values of
the CE efficiencies are also in agreement with first tentative 
results obtained from numerical simulations of the CE phase 
\citep{ricker+taam12-1}. 

However, we have no information whether these values 
hold for larger secondary masses.
While the decrease of $\alpha_{\mathrm{CE}}$ with increasing secondary masses 
proposed by \citet{demarcoetal11-1} seems unlikely \citep{zorotovicetal11-2}, 
the efficiencies may perhaps increase with secondary mass.  
The parameters of the only previously known PCEB with a G-type secondary, i.e., 
IK\,Peg ($\Porb=21.722\,d$, $\Mwd=1.19\Msun$, $\Msec=1.7\Msun$), indicate that 
we can probably not simply apply the constraints for PCEBs with M-dwarfs to 
larger secondary masses. As shown by e.g.,
\citet{davisetal10-1} and \citet{zorotovicetal10-1}, 
IK\,Peg is the only PCEB that requires 
additional energy sources to be at work during 
CE evolution. Despite the potential importance of IK\,Peg for our understanding 
of CE evolution, we cannot develop evolutionary theories based on just 
one system. Every new PCEB with a massive secondary therefore needs to be
carefully analyzed.

\begin{figure*}[t]
\begin{center}
\includegraphics[angle =0, width=0.83\textwidth]{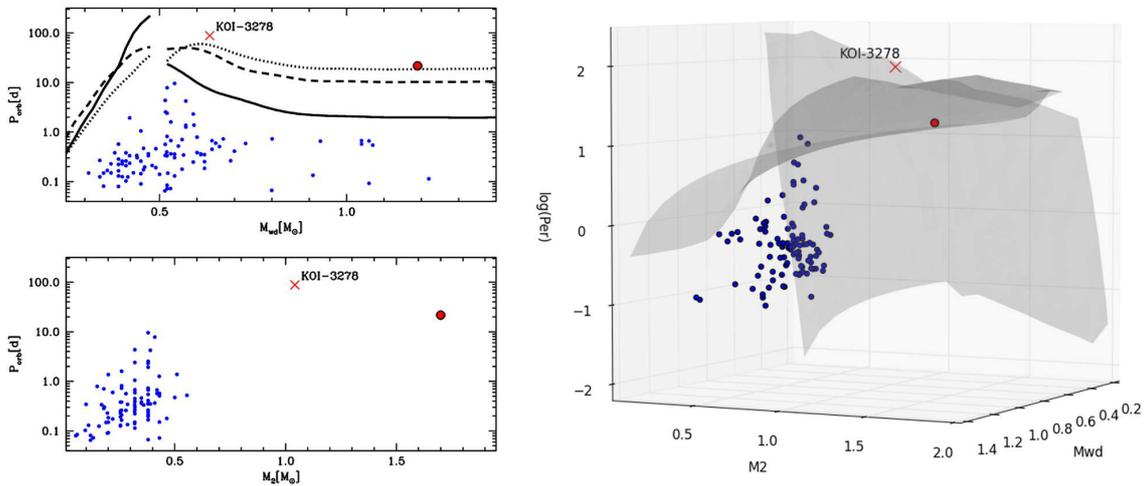}
\end{center}
\caption{\textit{Right:} Orbital period, secondary mass, and WD mass for the known PCEBs. The gray surfaces represent the maximum period of PCEBs
if orbital energy is the only energy source for expelling the envelope during the CE phase.
This maximum period is a function of \Msec\, and \Mwd. It increases with core mass for progenitors that fill their Roche lobe on the 
the first giant branch (i.e., for $\Mwd\lppr0.47\Msun$, gray surface in the back) and is generally long for AGB progenitors ($\Mwd\gppr0.51\Msun$, gray surface in the front).
KOI-3278 (red cross) and IK\,Peg (red point) are the only known PCEBs with solar-type secondary stars, and both systems are located above this limit. 
All WD+M-dwarf PCEBs (blue points) are well below the critical value. 
\textit{Left:} 2D projections of the right hand plot. The three black lines in the upper panel represent the maximum period for a fixed mass of 
the secondary star. The solid line is for 0.3\Msun (typical for the WD+M-dwarf PCEBs), the dashed line for 1.042\Msun (the companion in KOI-3278), 
and the dotted line for 1.7\Msun (the companion in IK\,Peg).}
\label{f-spdist}
\end{figure*}

\subsection{A note of caution for BSE users}

In their discovery paper, \citet{kruse+agol14-1} 
used the binary star evolution (BSE) code from \citet{hurleyetal02-1} and 
found a possible evolutionary path for KOI-3278 
assuming $\alpha_{\mathrm{CE}} = 0.3$  and $\lambda=0.2$. 
However, if we run BSE with the initial parameters obtained 
by \citet{kruse+agol14-1} with these values, 
the binary system does not survive the CE
phase. This discrepancy is easily explained by taking a closer look at the
evolution of BSE. In its original version, the code requested a fixed value for 
$\lambda$ as an input parameter. However, the code was frequently 
updated over the years
and a function to compute the value of $\lambda$ was included\footnote{the function 
called \textit{celamf} can be found in the file called \textit{zfuncs.f} and it was recently published in \citet{claeysetal14-1}.}. 
This change is not described in the README file and is not commented in the main code (\textit{bse.f}). 
However, digging into the code it becomes clear that in the current 
version the input parameter called ``\textit{lambda}'' 
represents the fraction of the recombination (ionization) energy 
that is included to compute the real value of $\lambda$, i.e., $\alpha_\mathrm{rec}$. 
If the user still wants to use a fixed value for $\lambda$, 
the input value must be negative 
(e.g., if one wants to use $\lambda=0.2$, the input parameter should be $-0.2$).

The result obtained by \citet{kruse+agol14-1} thus only shows that 
the evolutionary history of KOI-3278 can be understood if recombination energy
significantly contributes to expelling the envelope. Given the long orbital
period of KOI-3278, however, the crucial question is if it represents
the second system after IK\,Peg that {\em{requires}} additional energy 
sources to contribute during CE evolution. 

\subsection{Reconstructing KOI-3278}

Given the importance of understanding the evolution of PCEBs with massive
secondaries, we here properly reconstruct the evolution of KOI-3278. 
We use the BSE code from \citet{hurleyetal02-1} to identify possible 
progenitors of KOI-3278 and investigate whether additional energy sources are 
required to understand its evolutionary history. Our reconstruction algorithm 
is described in detail in \citet{zorotovicetal11-1}.

We first try to reconstruct the CE phase allowing the stellar parameters to vary within the 1-$\sigma$
uncertainties of the measured stellar parameters (as given by \citealt{kruse+agol14-1}) but without considering
additional energy sources. Interestingly, 
as in the case of IK\,Peg, we do not find possible progenitors for KOI-3278 
without violating energy conservation, i.e., for any possible progenitor of the 
WD, the CE could not have been expelled by the use of orbital 
energy alone. This represents an important result as KOI-3278 is only the 
second PCEB with a massive secondary star and, in contrast to all 
PCEBs with low-mass secondary stars, both these systems require
$\alpha_{\mathrm{rec}}>0$.  
Figure\,\ref{f-spdist} illustrates our finding. It shows the maximum orbital 
period that a system can have if the only energy source used to expel
the envelope is the orbital energy ($\alpha_{\mathrm{rec}}=0$ and 
$\alpha_{\mathrm{CE}}=1$) as a function of secondary mass, orbital period, 
and WD mass. The two systems with massive companions, 
KOI-3278 and IK\,Peg, are the only two that require extra energy sources. 
If a fraction of the recombination energy is assumed to 
contribute to expelling the envelope, the evolutionary history of both systems can be 
reconstructed. Assuming $\alpha_{\mathrm{CE}}=\alpha_{\mathrm{rec}}=0.25$
\citep{zorotovicetal10-1} we derive initial masses of $M_{1,i} = 2.450\Msun$ and 
$M_{2,i} = 1.034\Msun$ and an initial orbital period of $P_{orb,i} \sim 1300\,d$ for 
the progenitor of KOI-3278, which is similar to the values obtained by \citet{kruse+agol14-1}
for $\alpha_{\mathrm{CE}}=0.3$ and $\alpha_{\mathrm{rec}}=0.2$.

\section{Predicting the future of KOI-3278}

While reconstructing the evolution of KOI-3278 provides
new information about the efficiencies of CE evolution, 
the future of close binaries consisting of WDs and massive companions 
is equally important as these systems may either enter a second CE phase,
which may lead to a double WD (DWD), or start thermal-timescale mass
transfer. These configurations represent the two classical channels 
towards SN\,Ia, i.e. the double- and single-degenerate channel. 
Which of the two channels is taken by a given 
system depends on the timescale of nuclear evolution of the secondary and the 
timescale of orbital angular momentum loss until the secondary 
fills its Roche lobe. 

If the PCEB has a relatively long orbital period, the secondary is likely 
to evolve off the MS and fill its Roche lobe as a giant. This
configuration leads to dynamically unstable mass transfer if $q>q_{\mathrm{crit}}\sim1-1.5$.
It depends then on the CE efficiencies and the orbital
period at the onset of CE evolution if the system survives the CE phase or 
if the two stars merge. In the first case, a DWD is formed. The two WDs lose angular momentum
because of gravitational radiation and if this DWD has a total mass exceeding the
Chandrasekhar limit it may produce a SN\,Ia.

If, on the other hand, the PCEB has a short orbital period, angular momentum 
loss due to magnetic braking and gravitational radiation 
can cause the secondary to fill its Roche lobe while it is still on the MS. 
This will lead to thermal-timescale mass transfer (for $q\,\gappr\,1$) and the systems appear as a super-soft X-ray source, i.e., 
the mass-transfer rate is high enough to generate stable hydrogen burning on the surface of the WD,
allowing the WD mass to grow. It may explode as SN\,Ia if it reaches the Chandrasekhar limit.

We predict the future of KOI-3278 using BSE and the current system parameters as derived in \citet{kruse+agol14-1}. 
Given the age of the system and the mass of the secondary, the latter will evolve off the MS and fill its Roche lobe 
during the first giant branch, in $\sim9\,Gyr$. At that moment the orbital period will still be $\sim80\,d$, 
the secondary will have a total mass of $1.009\Msun$ with a core-mass of $0.332\Msun$. Given the mass ratio of the 
system, mass transfer will be dynamically unstable and lead to CE evolution. 
KOI-3278 will survive this second CE phase and form a DWD, a C/O WD plus a He WD,
for almost any value of the efficiencies (even if $\alpha_{\mathrm{rec}} = 0$, 
$\alpha_{\mathrm{CE}}$ needs to be only larger than $0.043$\footnote{This limit should be
slightly larger if we take into account that the WD that emerges from the CE phase is probably bloated
compared to a cool WD (as used in BSE).} to avoid a merger).
This makes KOI-3278 the first known progenitor of a DWD formed by two CEs. 

\begin{figure}[t]
\begin{center}
\includegraphics[angle =0, width=0.95\columnwidth]{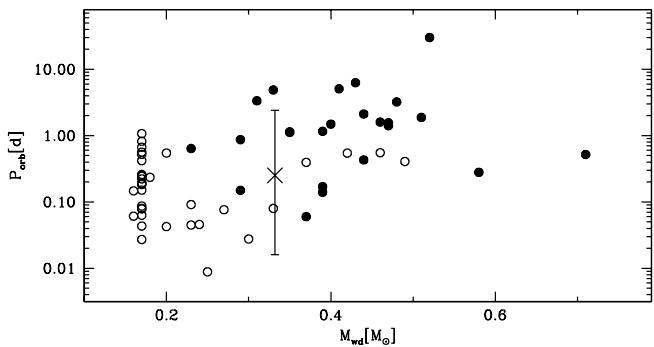}
\end{center}
\caption{Orbital period versus mass of the brighter WD (usually the younger one) for the DWDs found by the SPY survey (filled circles) 
and by the ELM survey (open circles). The cross marks the position KOI-3278 will take in $\sim9\,Gyr$. The error bar
has been calculated taking into account all possible values of the efficiencies during the second CE. }
\label{f-dwd}
\end{figure}

In Fig.\,\ref{f-dwd} we relate the future DWD parameters of KOI-3278 to 
the currently known sample of DWDs. The two surveys that have identified 
most of the currently known DWDs are the SPY survey \citep{napiwotzkietal03-1} and the 
ELM survey \citep{kilicetal10-1}. We compiled the DWDs from both surveys 
using the tables provided in
\citet{nelemansetal05-1} and \citet{brownetal12-1,brownetal13-1}. 
With an orbital period of $0.251^{+2.154}_{-0.234}\,d$ and a mass of $0.332\Msun$ 
for the WD that will form during the second CE phase, KOI-3278 will become 
a very typical DWD that will evolve towards shorter orbital 
periods driven by orbital angular momentum loss due to gravitational radiation.
Given the mass ratio, and depending on the strength of spin-orbit coupling, 
the binary may either become an AM\,CVn system or, more probably, merge 
\citep[see][ their figure\,1]{marshetal04-1}.

\section{Conclusion}

Understanding the evolution of the two known PCEBs containing a G-type 
secondary star requires additional sources of energy, such as
recombination energy, to contribute during CE evolution. 
This may indicate that a larger fraction of the
total available energy is used to expel the envelope. In other words,  
at least one of the efficiencies may increase with secondary mass.
If this can be confirmed, the population of 
PCEBs with F- and G-type secondaries will be dominated by long orbital period
systems ($\Porb\sim\,2-100\,d$) and most secondaries will evolve into giants 
before the second phase of mass transfer and may, such as KOI-3278, survive 
a second CE. This, finally, may imply that the double-degenerate channel towards
SN\,Ia is more likely to occur than the single-degenerate channel. 

\begin{acknowledgements}
We thank Fondecyt for their support under the grants 3130559 (MZ), 1141269 (MRS), and 3140585 (SGP).  
\end{acknowledgements}

\bibliographystyle{aa}

\end{document}